# Spin inversion in graphene spin valves by gate-tunable magnetic proximity effect at one-dimensional contacts

Jinsong Xu,[1] Simranjeet Singh,[1] Jyoti Katoch,[1] Guanzhong Wu,[1] Tiancong Zhu,[1] Igor Žutić,[2] Roland K. Kawakami[1*]

[1]Department of Physics, The Ohio State University, Columbus, OH 43210, USA

[2]Department of Physics, University at Buffalo, State University of New York, Buffalo, New York 14260, USA

* kawakami.15@osu.edu

**Graphene has remarkable opportunities for spintronics[1] due to its high mobility and long spin diffusion length, especially when encapsulated in hexagonal boron nitride (h-BN)[2-5]. Here, for the first time, we demonstrate gate-tunable spin transport in such encapsulated graphene-based spin valves with one-dimensional (1D) ferromagnetic edge contacts. An electrostatic backgate tunes the Fermi level of graphene to probe different energy levels of the spin-polarized density of states (DOS) of the 1D ferromagnetic contact, which interact through a magnetic proximity effect (MPE) that induces ferromagnetism in graphene. In contrast to conventional spin valves, where switching between high- and low-resistance configuration requires magnetization reversal by an applied magnetic field or a high-density spin-polarized current[6], we provide an alternative path with the gate-controlled spin inversion in graphene. The resulting tunable MPE employing a simple ferromagnetic metal holds promise for spintronic devices and to realize exotic topological states, from quantum spin Hall and quantum anomalous Hall effects[7,8], to Majorana fermions and skyrmions[9-11].**

Spintronics aims for information storage and logic operations by utilizing the electron's spin degree of freedom in addition to its charge. Beyond the three basic processes of spin injection, spin transport, and spin detection, it is crucial to explore new methods of spin manipulation[12-20] in order to develop novel architectures for spin-based logic. Graphene has become established as an excellent material for spin transport[1-5,21-23], and there recently have been important advances in the control and modulation of spin transport in graphene spin valves. In graphene with transition metal dichalcogenide (TMD) overlayers, the full modulation of spin currents was demonstrated using electrostatic gates to tune spin absorption into the TMD[17,19]. In graphene constrictions, the spin transport signal was controlled by gate-tunable quantum interference at low temperatures (< ~5 K)[14]. Alternatively, the MPE in graphene from a ferromagnetic insulator (FI) has been used to fully modulate spin currents by controlling the FI magnetization direction[18,20]. Beyond these experimental advances, Lazić *et al.* proposed a different type of MPE from simple ferromagnetic metals that can be used to invert the polarization of spin current as a function of electrostatic gate, but the gate-tunable MPE has yet to be realized[24,25].

In this paper, we experimentally demonstrate the gate-tunable MPE in encapsulated graphene spin valves with 1D ferromagnetic edge contacts and observe an inversion of the non-local spin signal as a function of electrostatic backgate voltage. To investigate the origin of the non-local spin signal inversion, we fabricate hybrid spin valves that employ both 1D contacts and traditional top contacts ("2D contacts") for spin injector and detector. By comparing the gate-dependent non-local spin signal for injector-detector combinations consisting of 1D-1D contacts vs. 1D-2D contacts, we are able to establish that the inversion of the non-local spin signal originates from an inversion of the effective spin polarization of the 1D contact as a function of gate voltage. This realizes the prediction by Lazić *et al.*[25] that a MPE associated with the spin-dependent DOS of a metallic ferromagnetic electrode could be used to modulate or even invert its effective spin polarization as a function of gate voltage. Unlike 2D contacts where the Fermi level of an adjacent graphene is strongly pinned[26-28], the Fermi level of graphene near a 1D contact can be tuned efficiently so that different energy levels of the ferromagnet's spin-polarized DOS can be accessed for gate-tunable MPE. This work provides a new method for manipulating both the magnitude and sign of the spin signal in graphene spin valves, and enables new device architectures for prospective graphene-based spin logic applications.

Figure 1a shows a scanning electron microscope (SEM) image of one of the measured devices (sample I), and Figure 1b is a schematic drawing of the device with three types of contacts: 1D transparent contacts, 1D tunneling contacts, or 2D tunneling contacts. The device consists of a h-BN/graphene/h-BN heterostructure on $SiO_2$(300 nm)/Si substrate (acting as backgate), contacted by nonmagnetic Cr/Au electrodes at two ends and ferromagnetic Co electrodes in between. Details of device fabrication are discussed in the Supplementary Information (SI), Section 1. Figure 1c shows a typical four-probe resistance of graphene as a function of gate voltage, $V_{gate}$ (the channel is 6 μm long and 1 μm wide). This yields a mobility of ~30,000 $cm^2$/Vs for holes and ~20,000 $cm^2$/Vs for electrons. To characterize the contacts, we perform three-terminal measurement of the differential contact resistance as a function of current bias at 20 K and 300 K (Figures 1d-1f). The 1D transparent contacts exhibit very little variation with either DC bias current or temperature, which is indicative of ohmic conduction (Fig. 1d). For contacts with 0.6 nm SrO barriers, both 1D (Fig. 1e) and 2D (Fig. 1f) contacts show strong bias-dependence of the contact resistance at 20 K, indicating non-ohmic conduction.

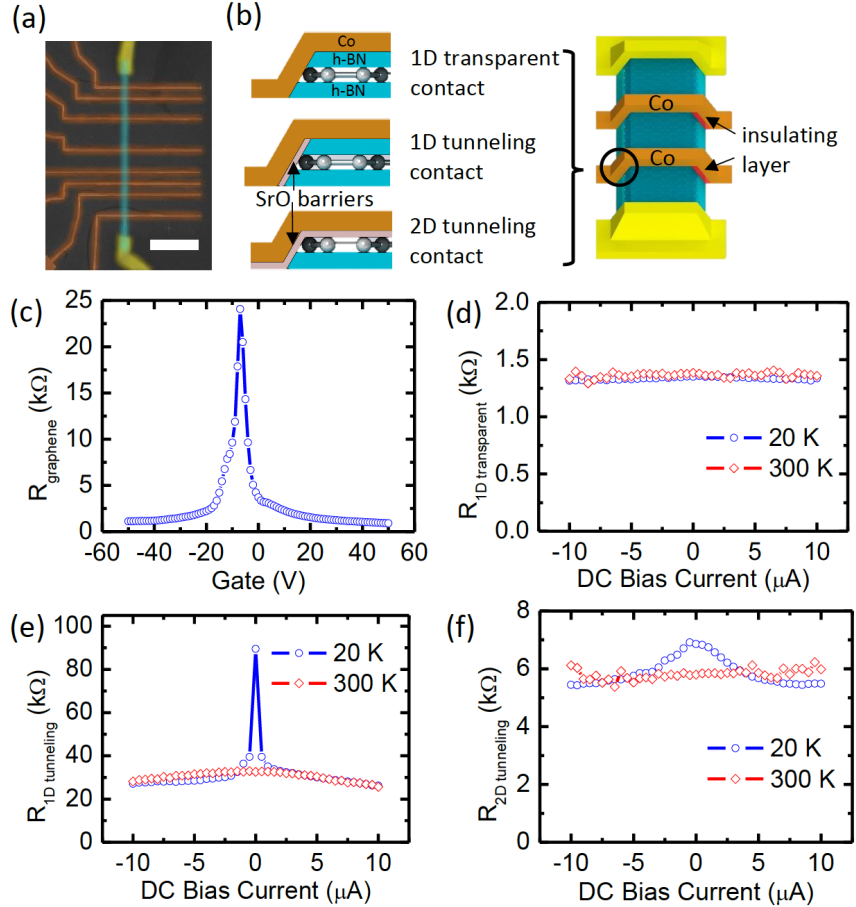

**Figure 1 | Electrical characterization of graphene and contact resistances. a,** Scanning electron microscope (SEM) image with false color of one of the measured devices (sample I) . The scale bar is 10 μm. **b,** Schematic of device and three different types of contacts. Co electrodes (brown color) form two 1D contacts with h-BN/graphene/h-BN heterostructure on two sides. The left side is 1D transparent/tunneling contact, and there is 3 nm thick SrO barrier (red color) isolating Co electrode from h-BN/graphene/h-BN on the right edge. So the current is only injecting from the left edge 1D contact. The yellow color is Cr/Au contact. **c,** Gate dependent graphene resistance (sample I). The channel length is 6 μm and the width is 1 μm. **d-f,** Typical bias dependence of the differential contact resistance for 1D transparent contacts, 1D tunneling contacts, and 2D tunneling contacts, respectively. All measurements are done with 0.1 μA AC excitation current.

Next, we investigate spin transport with 1D transparent contacts. All measurements are conducted in the non-local geometry (Fig. 2a inset) with lock-in detection and at 20 K. An injection current, $I_{INJ}$ (5 μA rms), is applied between the Co spin injector (E2) and nonmagnetic electrode (E1), and spin transport from E2 to Co spin detector (E3) is measured as a non-local voltage ($V_{NL}$) measured across E3 and nonmagnetic electrode (E4). The non-local resistance, $R_{NL}$, is defined as $R_{NL} = V_{NL}/I_{INJ}$. Figure 2a shows $R_{NL}$ as a function of external magnetic field along the Co electrode ($B_y$) at $V_{gate}$ = -40 V. The red (blue) circles are data for increasing (decreasing) $B_y$. At ~20 mT, an abrupt change in $R_{NL}$ is due to switching of the Co electrode magnetizations from parallel to antiparallel configuration. With further increasing magnetic field, $R_{NL}$ changes back to its original value as the Co electrode magnetizations align parallel. The spin transport is quantified by the change of non-local resistance $\Delta R_{NL} = R_{NL}$ (parallel) $- R_{NL}$ (antiparallel), which is ~0.2 Ω. Interestingly, as gate voltage is tuned to 0 V (Fig. 2b), the curves invert and $\Delta R_{NL}$ becomes negative.

To confirm that the non-local signal originates from spin transport, we perform in-plane non-local Hanle spin precession measurements[29,30] by applying an in-plane external magnetic field $B_x$ perpendicular to Co electrodes. The brown (purple) curves in Figure 2c and 2d are the in-plane Hanle curves for the parallel (antiparallel) configuration for $V_{gate}$ = -40 V and 0 V, respectively. This verifies the gate-dependent spin inversion with $\Delta R_{NL} > 0$ for $V_{gate}$ = -40 V and $\Delta R_{NL} < 0$ for $V_{gate}$ = 0 V. The extracted spin lifetimes and spin diffusion lengths are up to ~500 ps and ~10 μm (see SI section 7). In addition, we further verify that these signals do not originate from local Hall effects associated with magnetic fringe fields[31] or tunneling anisotropic magnetoresistance (see SI sections 8 & 9).

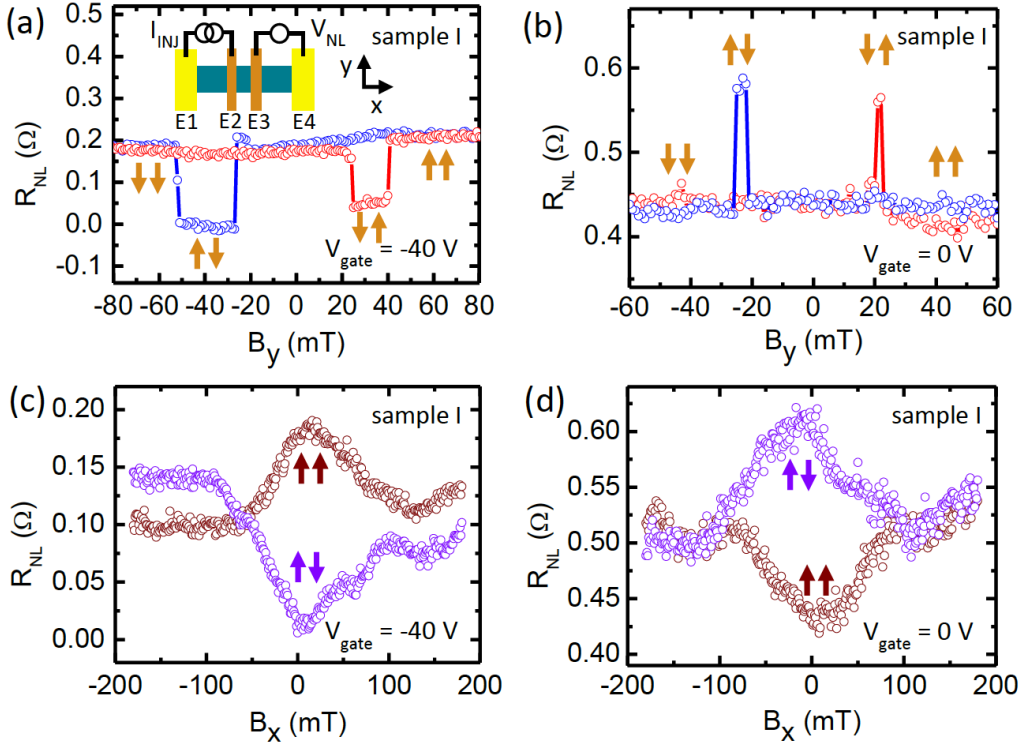

**Figure 2 | Spin inversion in graphene spin valves with 1D transparent contacts. a,b,** Non-local spin signal $R_{NL}$ (sample I) for gate voltages of $V_{gate}$ = -40 V and $V_{gate}$ = 0 V, respectively. The red (blue) curve is for increasing (decreasing) magnetic field. Inset: non-local spin transport geometry. **c,d,** Non-local Hanle measurement for gate voltages of $V_{gate}$ = -40 V and $V_{gate}$ = 0 V, respectively. The brown (purple) curve is for parallel (antiparallel) alignment of the injector and detector magnetizations. The lock-in measurements utilize a 5 μA excitation current.

The detailed gate dependence of the non-local spin signal (Fig. 3a) exhibits a sign reversal at $V_{gate}$ ~ -10 V. We note that gate-tunable spin inversion has been observed in most of the 1D edge contacted devices (10 out of 11 samples that exhibited spin signals). In addition, we observe substantial sample-to-sample variations including, in some cases, the polarity of the non-local spin signal changing more than once (see SI Section 2).

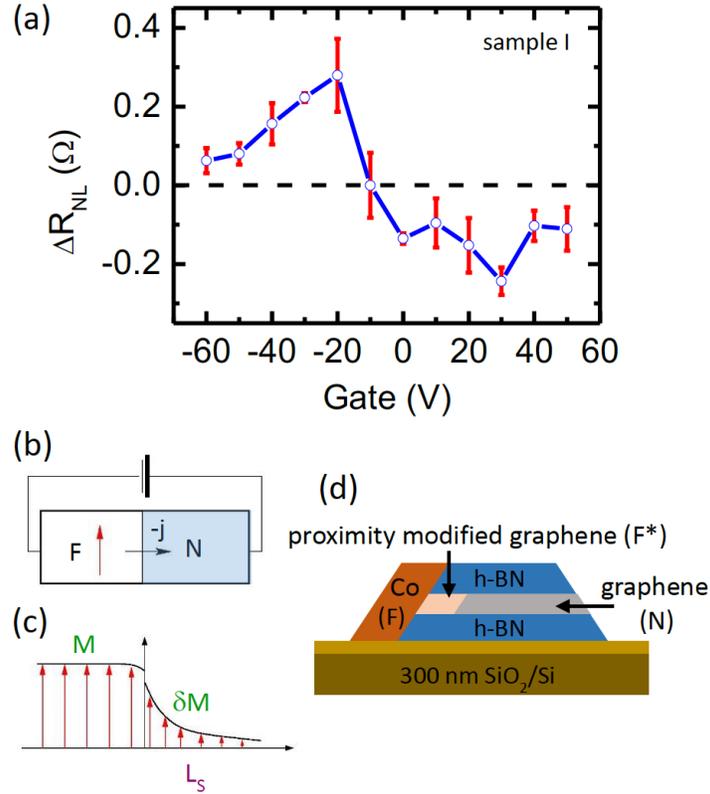

**Figure 3 | Gate-tunable magnetic proximity effect at 1D transparent contacts. a,** Gate dependence of the non-local spin signal, $\Delta R_{NL}$ (sample I). **b,** Schematic of bulk-like metallic ferromagnet/nonmagnet (F/N) junction (e.g. Co/Cu). **c,** Spin accumulation in bulk-like F/N junction. **d,** Schematic of the magnetic proximity effect for encapsulated graphene with a transparent 1D edge contact to a ferromagnetic electrode.

To understand the origin of the gate-tunable spin inversion, we first rule out quantum interference effects that invert spin signals in small constrictions (few hundred nm) and low temperatures (<5 K)[14] because our samples are larger (6 μm x 1 μm) and effects persist up to 75 K (see SI Section 3). Instead, the sign change of $\Delta R_{NL}$ is likely due to the spin-dependent DOS of the 1D ferromagnetic contact, which induces ferromagnetism in the graphene via MPE (see Figure 3d). In metallic ferromagnet/nonmagnet (F/N) bulk-like junctions, e.g. Co/Cu (Fig. 3b,c), the proximity-induced exchange splitting and magnetization, $M$, decay over a much shorter distance compared to the spin-diffusion length ($L_S$) in a nonmagnetic material, which determines the spin injection. Consequently, the description of spin injection usually completely neglects the equilibrium proximity effect in nonmagnetic materials. However, in atomically thin materials such as graphene, the magnetic proximity length exceeds their thickness. The wave functions from the metallic ferromagnet penetrate into graphene, directly polarizing its electronic structure at the Fermi level. Thus, describing spin injection for graphene spin valves with 1D edge contacts should include the MPE[24]. Instead of occurring at the F/N interface, the spin injection happens at the F* (proximity modified graphene)/N interface (Fig. 3d). Previously, evidence for such proximity modified graphene was observed in vertical Co/graphene/NiFe magnetic tunnel junctions, although the two-terminal geometry does not permit backgate tuning[32]. In addition, x-ray magnetic circular dichroism and angle resolved photoemission experiments directly confirm the magnetic proximity effect for graphene on ferromagnetic metals[33-35].

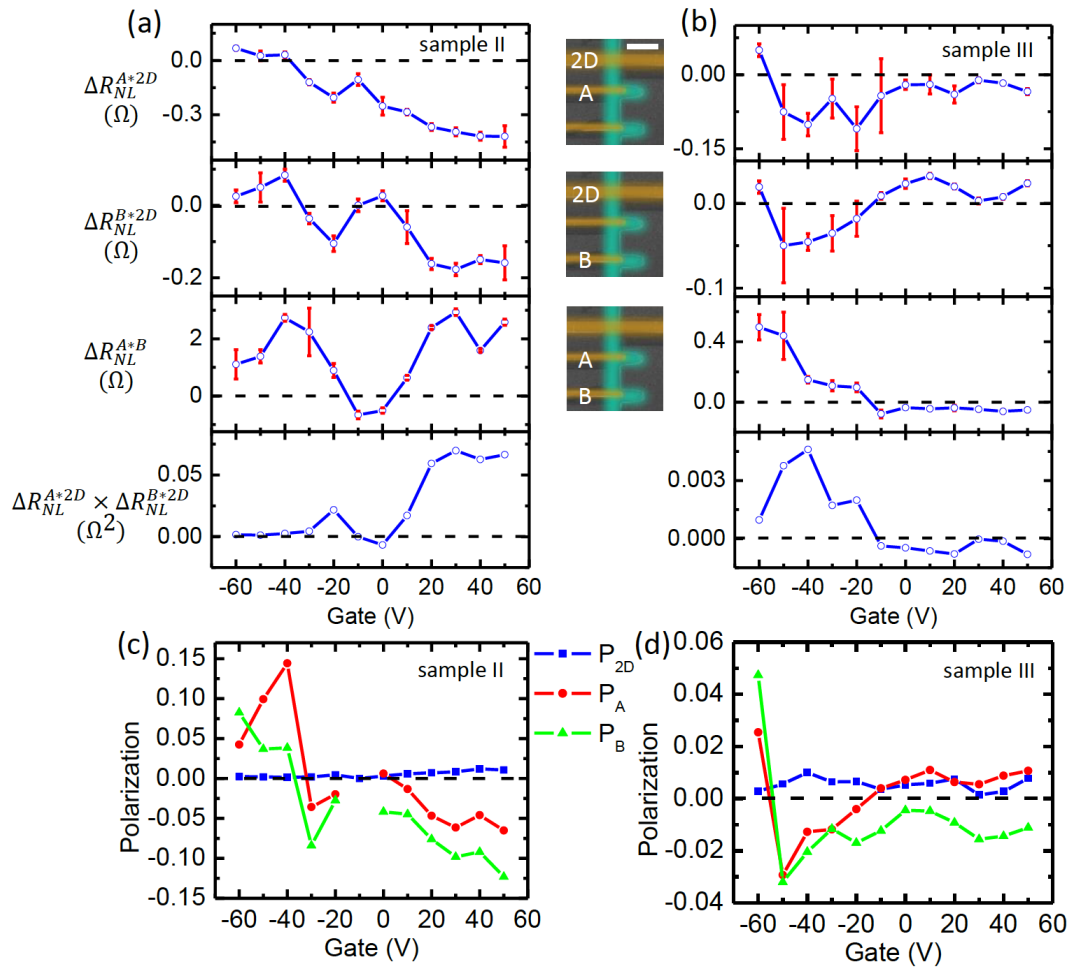

**Figure 4 | Spin inversion in hybrid graphene spin valves. a,b,** Gate dependence of $\Delta R_{NL}$ for two hybrid devices, sample II and sample III. The spin injector and detector are 1D contact A and 2D contact for the first row, 1D contact B and 2D contact for the second row, 1D contact A and 1D contact B for the third row, respectively, as shown in the central column of SEM images. The scale bar is 2 μm. The last row is the calculated product of the first two rows. **c,d,** Effective spin polarization of each contact for sample II and sample III.

To test whether the inversion of $\Delta R_{NL}$ originates from MPE, we need to demonstrate that the effective spin polarization of the 1D contact exhibits a sign reversal. We accomplish this by designing and testing hybrid spin valve devices with both 1D and 2D contacts. Because $\Delta R_{NL}$ is roughly proportional to the product of the injector and detector polarization ($\Delta R_{NL} \sim P_{inj}P_{det}$)[36,37], one can investigate spin valves with three ferromagnetic electrodes on the same graphene channel and perform three independent spin transport measurements using the three distinct pairings. Then, one can algebraically extract the effective spin polarization of each contact (see SI section 4 for details).

Figures 4a and 4b show the gate dependence of $\Delta R_{NL}$ for two hybrid devices (sample II and sample III), each consisting of two 1D transparent contacts ("A" and "B") and a 2D tunneling contact. The top row is $\Delta R_{NL}$ measured using 1D contact A and the 2D contact ($\Delta R_{NL}^{A*2D}$), while the second row is $\Delta R_{NL}$ measured using 1D contact B and the 2D contact ($\Delta R_{NL}^{B*2D}$). Here, the observed inversion of spin signal suggests a polarity reversal of the 1D contact because 2D contacts typically show weak gate dependence (see SI section 5). The third row is $\Delta R_{NL}$ measured using 1D contact A and 1D contact B ($\Delta R_{NL}^{A*B}$), while the last row ($\Delta R_{NL}^{A*2D} \times \Delta R_{NL}^{B*2D}$) is the calculated product of the first two rows. By comparing $\Delta R_{NL}^{A*B}$ with $\Delta R_{NL}^{A*2D} \times \Delta R_{NL}^{B*2D}$, we notice that they are very similar to each other, including both the sign and the trend. Because $\Delta R_{NL}^{A*B} \sim P_A P_B$ and $\Delta R_{NL}^{A*2D} \times \Delta R_{NL}^{B*2D} \sim P_A P_B (P_{2D})^2$, the similarity of the two curves suggests that $P_{2D}$ has relative little gate dependence, as expected. For quantitative analysis, we utilize the Takahashi-Maekawa model[36] of non-local spin transport and solve for the effective spin polarizations $P_A$, $P_B$, and $P_{2D}$ in terms of the experimental $\Delta R_{NL}^{A*2D}$, $\Delta R_{NL}^{B*2D}$, and $\Delta R_{NL}^{A*B}$ data (see SI section 4 for details). The resulting gate-dependent effective spin polarizations are shown in Figures 4c and 4d for samples II and III, respectively. Although the two samples have different gate dependences, two important trends are observed in both samples. First, as expected, the effective polarization of the 2D contact, $P_{2D}$, exhibits relatively little variation with gate voltage. Second, the 1D contacts exhibit polarity inversion as a function of gate voltage. Because the gate-dependent inversion of $\Delta R_{NL}$ occurs due to polarity inversion of the 1D contacts, this strongly supports our hypothesis that the spin inversion originates from gate-tunable MPE.

Theoretically, this strong gate dependence of the effective spin polarization of the 1D contact comes from the spin-dependent DOS of the 1D interface between graphene and Co as a result of the proximity effect. Lazić *et al.* reported that the band structure of graphene becomes spin-dependent due to proximity effect to ferromagnetic metals[24,25]. As a result, one can tune the polarity of spin polarization at the interface through adjusting the Fermi level in graphene by gate voltage. In conventional graphene spin valves with 2D contacts, the Fermi level of graphene underneath the ferromagnetic electrode is strongly pinned and cannot be tuned effectively by gate voltage. This is supported by first-principles calculations[25,38], which show that for a 2D contact (e.g. Co/h-BN/graphene), the graphene becomes strongly n-doped: its Dirac cone is shifted ~ -0.5 eV below the Fermi level. A simple electrostatic model[25] predicts that the small graphene DOS is required for effective gating. While this could be possible in 2D contacts by shifting the low-DOS region of the Dirac cone close to the Fermi level, it would require a very large gate-induced electric field, more than an order of magnitude larger than ~0.2 V/nm from our experiments (~60 V voltage applied on 300 nm $SiO_2$ + ~20 nm h-BN dielectric material). In contrast, our spin valves with 1D contacts form less aggressive contacts with Co which have a much smaller effective n-doping of graphene and the shift of the Dirac cone below the Fermi level than what is calculated for 2D contacts[25,38]. Consequently, in 1D contacts graphene's low-DOS region is attained by a much smaller gate voltage, ensuring that different spin-dependent states are readily accessible by gate tuning. Furthermore, complementary studies of nonlocal spin signal in standard graphene spin valves exhibit spin inversion as a function of bias current [23], and a comparison of different barrier materials (i.e. h-BN, MgO)[39] provides evidence for a DOS of Co that reverses spin polarization as a function of energy. Thus, the 1D contacts provide the two necessary ingredients for achieving gate-dependent inversion of the magnetic proximity effect[25], namely: (1) having the Dirac point of graphene in the F contact region to be near the Fermi level, where the low DOS in graphene at the Fermi level enables substantial changes in P, and (2) having a F DOS with reversal of spin polarization as a function of energy.

We further examine the MPE by inserting a 0.6 nm SrO tunnel barrier at the 1D contact and again observe gate-tunable spin inversion (see SI Section 6 for details). This demonstrates that the MPE can extend across a tunnel barrier, as predicted theoretically[25].

In conclusion, we demonstrate non-local spin transport in h-BN/graphene/h-BN heterostructures with 1D ferromagnetic contacts, and the non-local spin signal can be tuned by gate voltage effectively and even changes polarity. By designing and testing hybrid spin valve devices with both 1D and 2D contacts, we demonstrate this intriguing behavior originates from the gate-tunable MPE at the 1D contact interface. These results pave the way for future development of graphene spintronics. Gate-controlled spin polarity may overcome the usual need for an applied magnetic field and a magnetization reversal to implement the graphene-based spin logic[40]. On the other hand, the interplay between the tunable MPE and the surface inversion asymmetry[1,6] in ferromagnet/h-BN/graphene-based heterostructures could support topologically nontrivial states, from quantum spin Hall and quantum anomalous effects[7,8], to Majorana fermions and skyrmions[9-11].

**Acknowledgment**

We thank Wei Han and Kirill Belashchenko for helpful discussions. Funding for this research was provided by the Center for Emergent Materials: an NSF MRSEC under award number DMR-1420451.

We also acknowledge support from C-SPIN, one of the six SRC STARnet Centers, sponsored by MARCO and DARPA. Igor Žutić is sponsored by US ONR 000141712793 and NSF-ECCS 1508873.

**Author contributions**

R. K. K. and J. X. conceived the study and J. X. performed the experiments. S. S., J. K., G. W., and T. Z. contributed to the device fabrication and heterostructure stacking infrastructure. I. Z. led the theoretical analysis. All authors contributed to the analysis and interpretation of the results and preparation of the manuscript.

# Supplementary Information for:

# Spin inversion in graphene spin valves by gate-tunable magnetic proximity effect at one-dimensional contacts

Jinsong Xu,[1] Simranjeet Singh,[1] Jyoti Katoch,[1] Guanzhong Wu,[1] Tiancong Zhu,[1] Igor Žutić,[2] Roland K. Kawakami[1]

[1]*Department of Physics, The Ohio State University, Columbus, OH 43210, USA*

[2]*Department of Physics, University at Buffalo, State University of New York, Buffalo, New York 14260, USA*

1. **Device fabrication**

The general procedure for fabricating h-BN/graphene/h-BN heterostructures is shown in Figure S1. First, we mount ~2 mm thick polydimethylsiloxane (PDMS) on a glass slide and cover it with a thin film of polycarbonate (PC). This PC/PDMS stamp is used to pick up top the h-BN flake from an $SiO_2$/Si substrate. The top h-BN flake is then aligned and brought into contact with graphene on an $SiO_2$/Si substrate to pick up graphene with this top h-BN flake. Then the whole stack is aligned and brought into contact with the bottom h-BN flake on an $SiO_2$/Si substrate. After contact, the PC film is cut from the glass slide and the entire PC/h-BN/graphene/h-BN combination remains on $SiO_2$/Si substrate. The PC film is then dissolved in chloroform. After that, the transferred h-BN/graphene/h-BN heterostructure is cleaned of polymer residue by annealing at 350 °C in ultra-high vacuum (UHV) for 1 hour. Then the h-BN/graphene/h-BN graphene heterostructure is patterned by e-beam lithography with PMMA resist and etched by low-power inductively coupled plasma reactive ion etch (ICP-RIE) to get the desired geometry. This process is followed by another annealing step in UHV to remove PMMA residue. Subsequently, we use two steps of e-beam lithography with MMA/PMMA bilayer resist to fabricate electrodes. In the first step, Au electrodes (70 nm) are deposited on the h-BN/graphene/h-BN heterostructure using an e-beam source and a 5 nm Cr underlayer for adhesion. In the second step, Co (60 nm) electrodes are directly deposited in an MBE chamber for one-dimensional (1D) transparent contacts. For tunnel barrier contacts, Co electrodes with SrO tunnel barriers are deposited using angle evaporation with polar angle of 0° for the SrO masking layer (3 nm), 10° for the SrO tunnel barrier (0.6 nm), and 6° for the Co electrode (60 nm). As shown in Figure S2, there are a total of four different device geometries: (1) strip shape with 1D contact, (2) Hall bar shape with 1D contact, (3) 2D contact and (4) 1D combining with 2D contact. For the strip shape device, on the right edge of the heterostructure, a 3 nm SrO barrier is deposited before Co deposition to form insulating layer to block any conduction. For 2D contact and 1D combining with 2D contact device, the top h-BN, before transfer, is etched with several slits which are used for 2D contact deposition. That is, there is no top h-BN in the red color Co electrode region in Figure S2c and S2d.

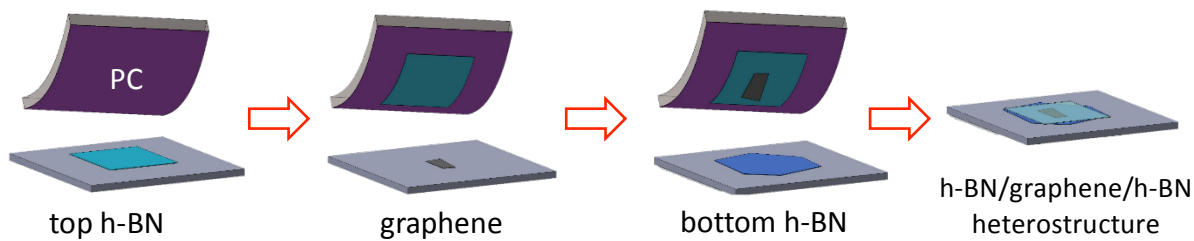

Figure S1. Schematic of transfer process

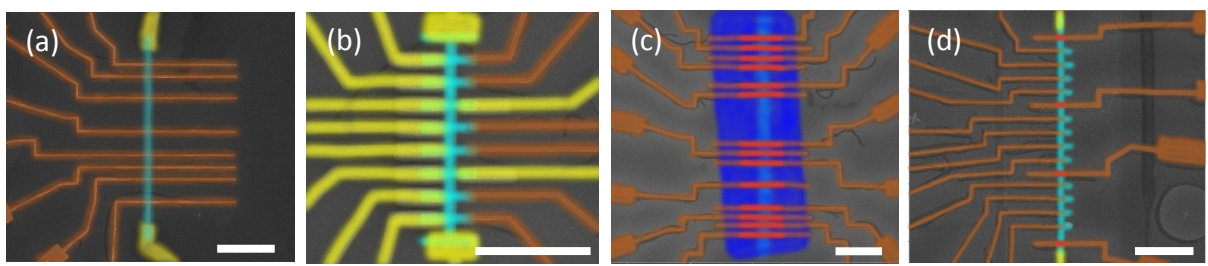

Figure S2. Scanning electron microscope images (false color) of four different device geometries. a, strip shape with 1D contact. b, Hall bar shape with 1D contact. c, 2D contact and d, 1D combining with 2D contact. Light blue is h-BN/graphene/h-BN heterostructure, yellow is Cr/Au electrode, brown is Co electrode and blue is h-BN. The scale is 10 μm.

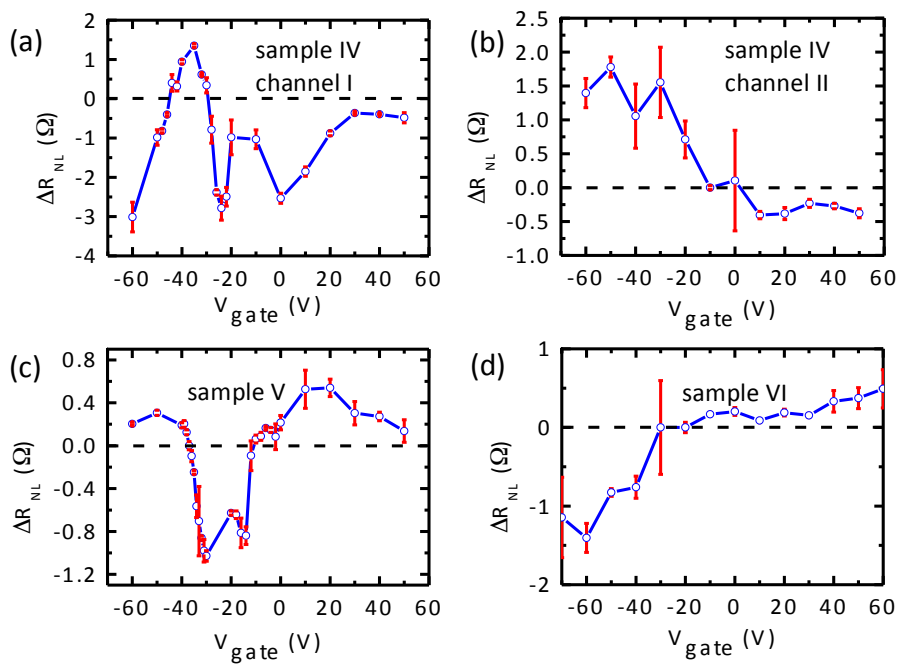

Figure S3. Additional data from different samples

## 2. Additional data for different samples

The change in polarity of $\varDelta R_{NL}$ as a function of gate voltage is observed in multiple samples with different geometries, which is shown in Figure S3. Figure S3a and S3b are data from two different pairs of electrodes in sample IV and Figure S3c is data from sample V (both sample IV and V have the geometry shown in Figure S2a). Figure S3d is data from a Hall bar device, sample VI, shown in Figure S2b.

## 3. Temperature dependence of $\varDelta R_{NL}$ for different backgate voltage

Compared to conventional graphene spin valves, there is stronger temperature dependence of $\varDelta R_{NL}$ for 1D contact devices. Figure S4 shows the temperature dependence of $\varDelta R_{NL}$ for different backgate voltage. The negative $\varDelta R_{NL}$ is observable up to 75 K and vanishes at 100 K for $V_{gate}$ = 0 V, 20 V and 40 V, while positive $\varDelta R_{NL}$ persists to 200 K and disappears at room temperature for $V_{gate}$ = -20 V, -30 V and -50 V.

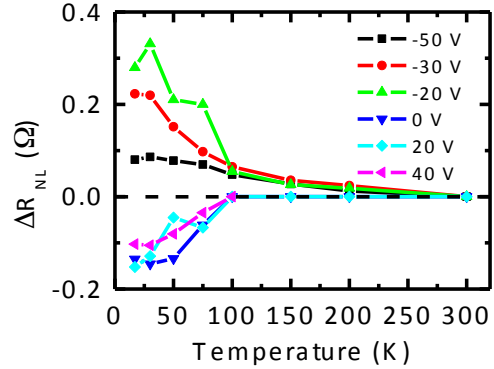

Figure S4. Temperature dependence of $\Delta R_{NL}$ for different gate (sample I)

## 4. Extracting the effective spin polarizations of contacts

For quantitative analysis we estimate the polarization of each contact using the model developed by Takahashi and Maekawa[1].

$$\Delta R_{NL} = 2R_N e^{-L/\lambda} \frac{\prod_{i=1}^{2}\left(\frac{P_{\Sigma i} R_{Ci}}{R_N}{1-P_{\Sigma i}^2} + \frac{P_\sigma^F R_F}{R_N}{1-P_\sigma^{F^2}}\right)}{\prod_{i=1}^{2}\left(1+\frac{\frac{2R_{Ci}}{R_N}}{1-P_{\Sigma i}^2}+\frac{\frac{2R_F}{R_N}}{1-P_\sigma^{F^2}}\right)-e^{-2L/\lambda}}, \quad (1)$$

where $R_N = \frac{\rho_G \lambda}{w}, R_F = \frac{\rho_F \lambda_F}{A_J}$ represents the spin resistance of graphene and Co, $\lambda$ and $\lambda_F$ are spin diffusion length of graphene and Co, $L$ and $w$ is the length and width of graphene (for sample II and III, we use $\lambda = 7.5\ \mu m$, $L_{A*2D} = 2.5\ \mu m$, $L_{B*2D} = 5\ \mu m$, $L_{A*B} = 2.5\ \mu m$ and $w = 1\ \mu m$), $A_J$ is Co electrode cross section, $R_C$ is contact resistance, $P_\sigma^F$ is Co spin polarization and $P_\Sigma$ is contact spin polarization, which we consider to be gate-dependent. Because in our samples $P_\Sigma$ is usually less than 0.1 and $R_F$ is also much smaller than $R_C$, equation (1) can be simplified to

$$\Delta R_{NL} = P_{\Sigma 1} P_{\Sigma 2} f_{12}(R_{C1}, R_{C2}, R_N, L, \lambda), \tag{2}$$

where
$$f_{12}(R_{C1}, R_{C2}, R_N, L, \lambda) = \frac{2R_N e^{-L/\lambda} \frac{R_{C1}}{R_N} \frac{R_{C2}}{R_N}}{(1+\frac{2R_{C1}}{R_N})(1+\frac{2R_{C2}}{R_N}) - e^{-2L/\lambda}}. \tag{3}$$

From equation (2) we can see that the spin signal is the product of spin polarization of spin injector and detector multiplied by a factor $f_{12}(R_{C1}, R_{C2}, R_N, L, \lambda)$ determined by $R_C$, $R_N$, $L$ and $\lambda$. Since the polarization of 2D contact does not change its polarity with gate voltage, to adopt a convention that its spin polarization is positive (i.e. we can only determine relative polarizations, so we must adopt a sign convention). Then we have

$$\begin{cases} P_{2D} = \sqrt{\frac{\Delta R_{NL}^{A*2D} \Delta R_{NL}^{B*2D}}{\Delta R_{NL}^{A*B}} \frac{f_{A*B}}{f_{A*2D} f_{B*2D}}} \\ P_A = \frac{\Delta R_{NL}^{A*2D}}{f_{A*2D}} \frac{1}{P_{2D}} \\ P_B = \frac{\Delta R_{NL}^{B*2D}}{f_{B*2D}} \frac{1}{P_{2D}} \end{cases} \tag{4}$$

Based on the three measured gate dependent non-local spin signal $\Delta R_{NL}^{A*2D}$, $\Delta R_{NL}^{B*2D}$ and $\Delta R_{NL}^{A*B}$, the contact resistance and graphene resistance at each gate voltage, we calculate the effective spin polarization of each contact using equation (4). The extracted effective spin polarization of each contact for sample II and III are shown in main text Figure 4c and 4d.

5. **Backgate dependence of encapsulated spin valve devices with 2D tunneling contacts**

Figure S5 shows the backgate dependence of the non-local spin signal $\Delta R_{NL}$ for an encapsulated spin valve device with 2D tunneling contacts (shown in Figure S2c). The device exhibits weak backgate dependence of $\Delta R_{NL}$ from 20 K up to room temperature.

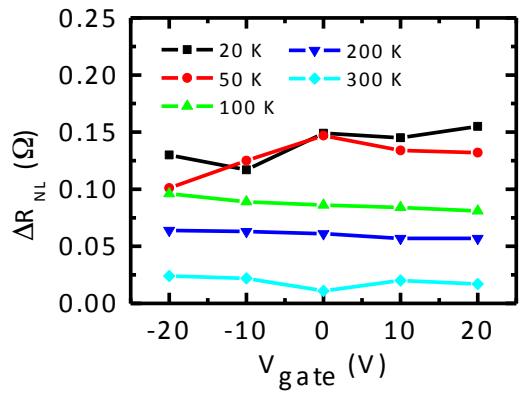

Figure S5. Backgate dependence of encapsulated spin valve device with 2D contacts at different temperature (sample VII)

## 6. Gate tunable proximity effect with 1D tunneling contact

Our results also suggest that the tunable magnetic proximity effect can exist in the presence of 1D contacts with a tunnel barrier. We measured h-BN/graphene/h-BN 1D contact devices with 0.6 nm SrO tunneling barriers. Figures S6a and S6b are the non-local magnetoresistance (MR) curves at $V_{gate}$ = - 40 V and $V_{gate}$ = 40 V. It is clear that the polarity of $\Delta R_{NL}$ still changes, even with 0.6 nm SrO tunneling barriers. This agrees with the theory prediction that the magnetic proximity effect can extend across a tunnel barrier[2].

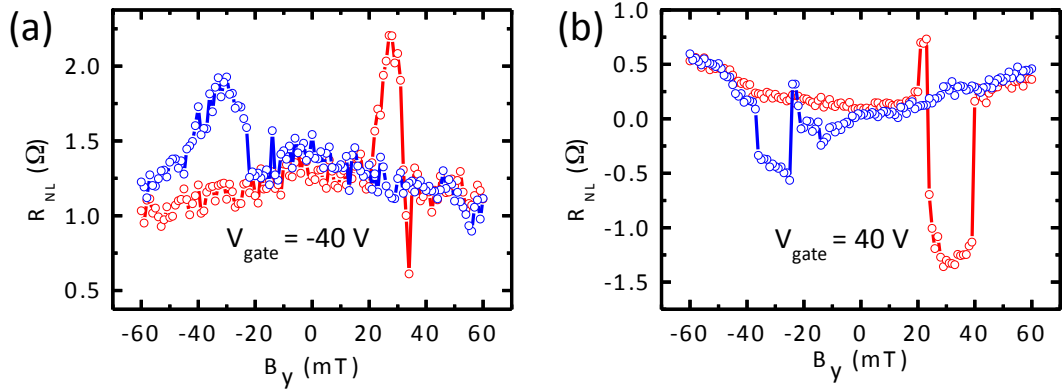

Figure S6. Non-local MR curves for graphene spin valve device with 1D tunneling contact. a, $V_{gate}$ = - 40 V and b, $V_{gate}$ = 40 V. The red (blue) curve is for increasing (decreasing) magnetic field. (sample VIII)

## 7. Spin lifetime and diffusion length

To extract spin lifetime and diffusion length, we perform the analysis on the raw data as well as symmetrized data because the curves show significant asymmetry as a function of Bx. We speculate shi asymmetry could be due to the non-collinearity between the spin orientation of the 1D interfaces at spin injector and detector. Figures S7a and S7b show the raw and symmetrized data for $V_{gate}$ = -40 V, respectively. For raw data in Figure S7a, we fitted with both standard model[3] (red curve) and modified model considering asymmetric component (blue curve). The origin of the asymmetry is not known but one possibility is the presence of a relative angle between the directions of the effective polarizations of the injector and detector. The extracted spin lifetime and diffusion length are 252 ± 62 ps (208 ± 12 ps for asymmetric fitting) and 8.6 ± 1.6 μm (7.7 ± 0.4 μm for asymmetric fitting). The extracted spin lifetime and diffusion length are 252 ± 12 ps and 8.6 ± 0.3 μm for symmetrized data in Figure S7b. Figures S7c and S7d show the raw and symmetrized data for $V_{gate}$ = 0 V, respectively. The extracted spin lifetime and diffusion length are 515 ± 122 ps (515 ± 87 ps for symmetrized) and 10.1 ± 1.3 μm (10.0 ± 1.0 μm for symmetrized).

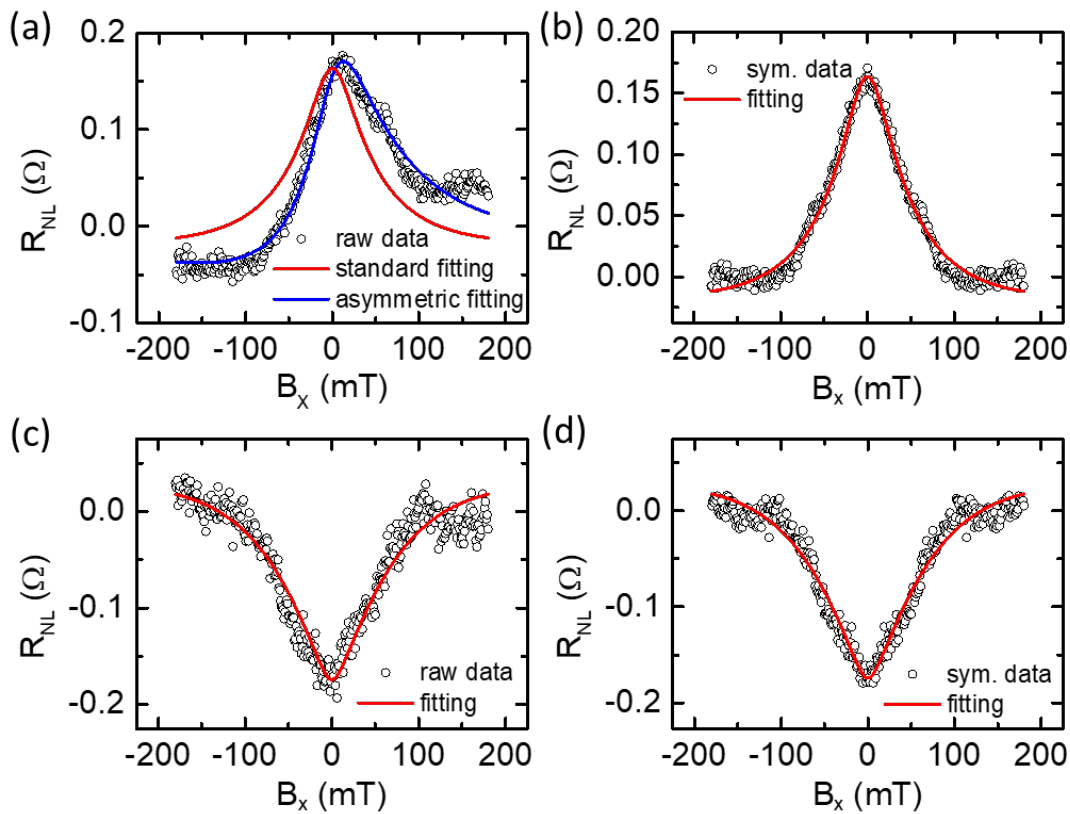

Figure S7. Spin lifetime and diffusion length of Sample I shown in main text. Hanle fitting for data shown in main text Figure 2c: a. raw data fitted with standard model (red curve) and modified model considering asymmetric component (blue curve) and b. symmetrized data fitted with standard model. Hanle fitting for data shown in main text Figure 2d: c. raw data and d. symmetrized data fitted with standard model

## 8. Comparison of local Hall effect and our spin transport

One concern is the local Hall effect from the fringe fields of Co electrodes, which may cause some artifacts. Recently published work of B. Karpiak *et al.*[4], suggests that in 1D edge ferromagnet/graphene contacts using a geometry and materials similar to ours, the observed results are dominated by such magnetic stray fields, not the magnetic proximity effects. To compare local Hall effect from fringe fields and our spin transport results, we fabricated a device (sample IX) with the geometry used in B. Karpiak *et al.* to purposely generate magnetic fringe fields and local Hall effects. Here, the Co electrodes terminate on top of the graphene channel instead of crossing the entire graphene, as shown in Figure S8a. Figure S8b is the gate dependent graphene channel resistance, of which the Dirac point is around $V_{gate}$ = -35 V.

Figure S8c is the Hall resistance $R_{xy}$ as a function of magnetic field $B_z$ at $V_{gate}$ = -10 V. The slope $d(R_{xy})/d(B_z) = 1/qe$ of this curve is the Hall coefficient $R_H$, which is summarized in Figure S8d for different gate voltage. The Hall coefficient changes sign at Dirac point. Figure S8e shows magnetic field $B_y$ dependent non-local resistance $R_{NL}$ at $V_{gate}$ = -40 V and it has a square shape 'hysteresis loop' which is due to the switch of fringe fields from a single Co electrode, while our spin valves data (main text Figure 2 and SI Figure S6) have two jumps due to spin transport from magnetization switches of two Co electrodes. Figure S8f shows magnetic field $B_y$ dependent non-local resistance $R_{NL}$ at $V_{gate}$ = -20 V, which has opposite sign to $R_{NL}$ at $V_{gate}$ = -40 V. This is expected from local Hall effect. Figure S8g summarizes the change of non-local resistance $\Delta R_{NL}$ as a function of gate voltage (red dots). It has the same trend to the Hall coefficient $R_H$ (blue curve). This strongly indicates the signal $\Delta R_{NL}$ is due to local Hall effect from the fringe fields of a single Co electrode. In contrast, our spin valve devices have totally different gate dependence. Figure S9 is the gate dependent graphene resistance for different samples. Different channels on the same sample have the same Dirac point (SI Figure S9a, S9b, S9c-d), indicating the samples are uniform. However, the gate dependence of $\Delta R_{NL}$ for different spin transport channels on the same sample (main text Figure 4a and 4b, SI Figure S3a and S3b) is different and $\Delta R_{NL}$ for different channels can have different signs at the same gate voltage, i.e. the same carrier type and density because the sample is uniform. And the sign change of $\Delta R_{NL}$ in some samples can even happen more than once (SI Figure S3a and S3c). Furthermore, the voltage signal from fringe fields should have a linear dependence on current, i.e. the non-local resistance signal $\Delta R_{NL}$ should be current independent as shown in Figure S8h, which is very different from our spin valve data as shown in Figure S10. In addition, while the spin signal $\Delta R_{NL}$ from MPE has a strong temperature dependence (SI Figure S4), $\Delta R_{NL}$ due to the local Hall effect from the fringe fields of Co electrode has weak temperature dependence and is observed up to room temperature, as shown in Figure S11. All these difference between local Hall effect from fringe fields and our spin valves data strongly indicate our spin valves signal is not from local Hall effect from fringe fields, but from spin transport.

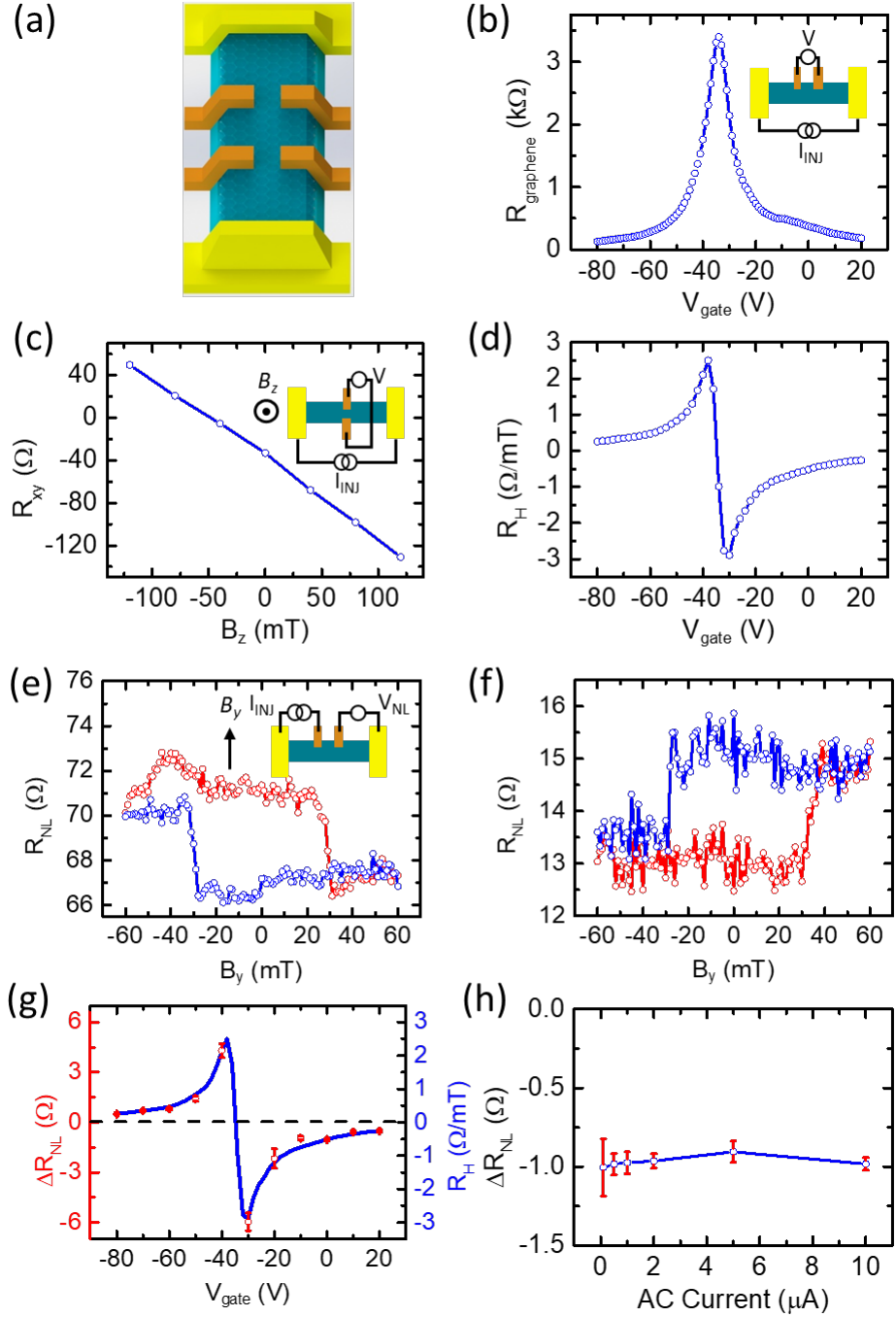

Figure S8. Low temperature (T = 20 K) transport measurement on sample IX with fringe fields from the end of Co electrodes. a, schematic drawing of sample IX, with Co electrodes end on graphene channel generating fringe fields. b, gate dependent graphene channel resistance. c. Hall resistance $R_{xy}$ as a function of magnetic field $B_z$ at $V_{gate}$ = -10 V. d. gate dependent Hall co-efficient $R_H$. e. and f., $R_{NL}$ as a function of magnetic field $B_y$ at $V_{gate}$ = -40 V and -20V. The red (blue) curve is for increasing (decreasing) magnetic field. g. gate dependent $\Delta R_{NL}$ (red dots) compared to Hall co-efficient $R_H$ (blue curve). h., $\Delta R_{NL}$ as a function of AC current for sample IX. The insets in Figure b, c, and e are measurement diagram.

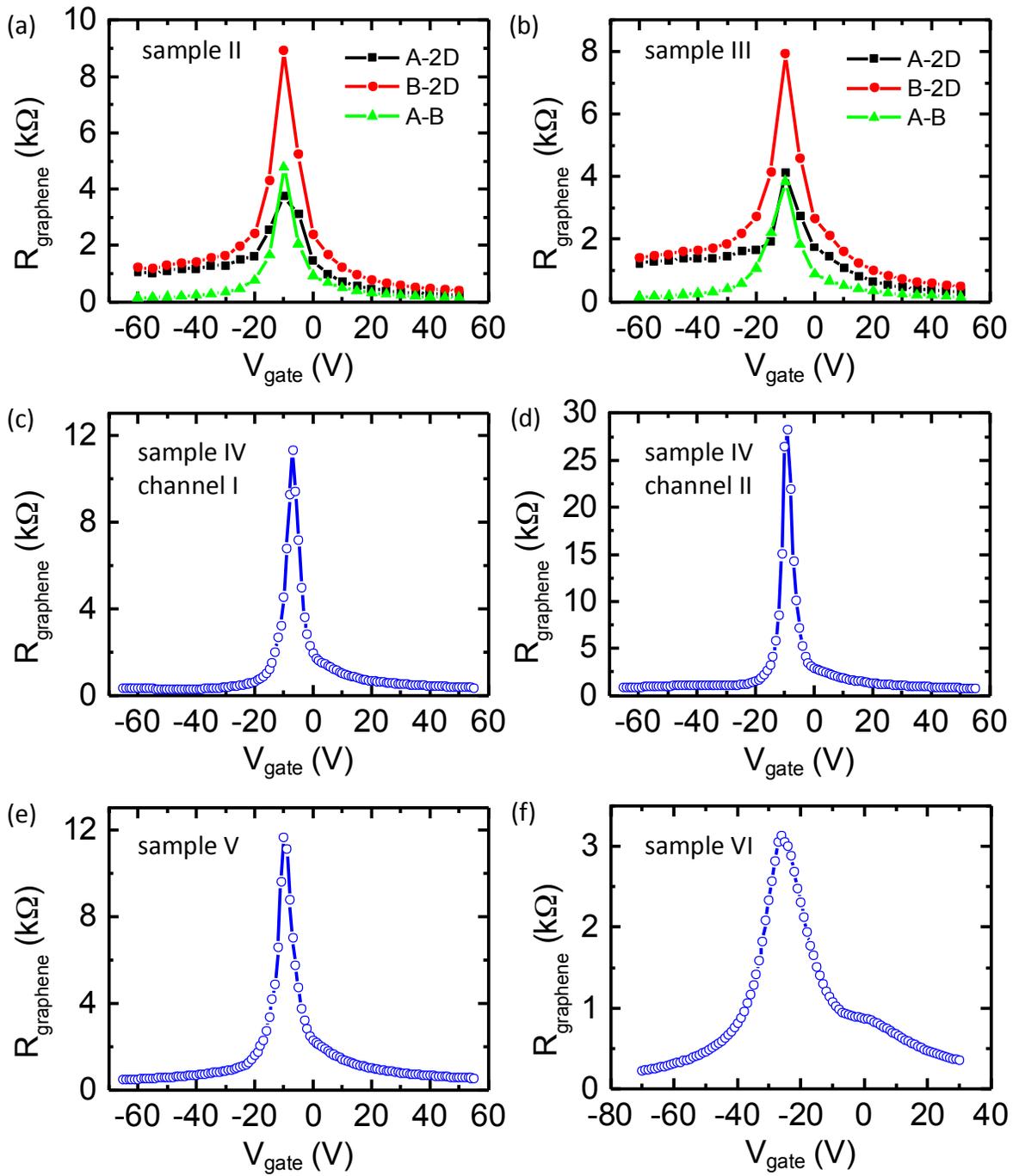

Figure S9. Gate dependent graphene channel resistance for different samples

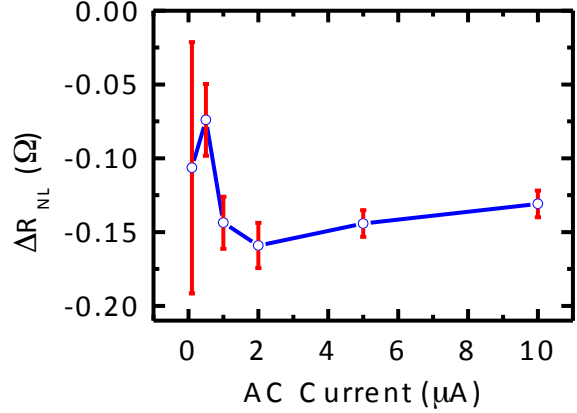

Figure S10. AC current dependent non-local resistance $\Delta R_{NL}$ from graphene spin valves sample I.

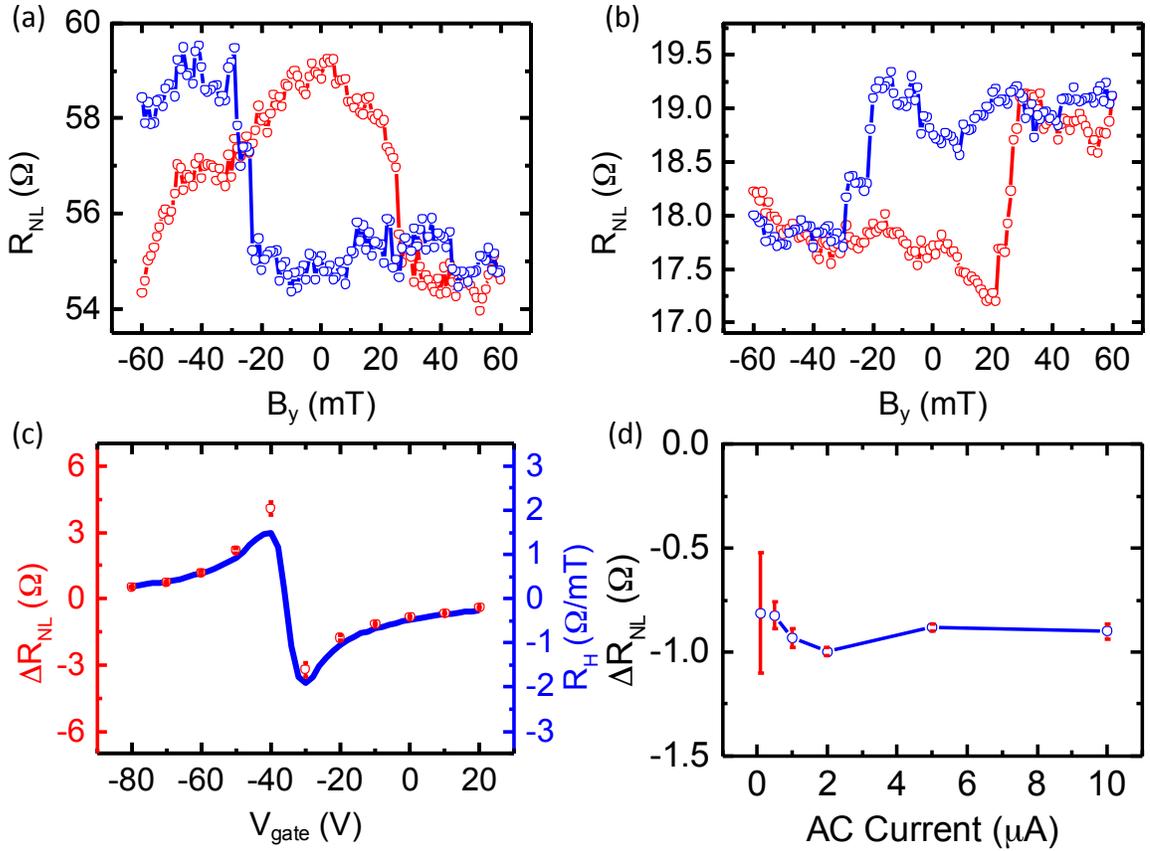

Figure S11. Room temperature (T = 300 K) transport measurement on sample IX with fringe fields from the end of Co electrodes. a. and b., $R_{NL}$ as a function of magnetic field $B_y$ at $V_{gate}$ = -40 V and -20V. The red (blue) curve is for increasing (decreasing) magnetic field. c. gate dependent $\Delta R_{NL}$ (red dots) compared to Hall co-efficient $R_H$ (blue curve). d., $\Delta R_{NL}$ as a function of AC current for sample IX.

## 9. Tunneling anisotropic magnetoresistance

Another concern is tunneling anisotropic magnetoresistance (TAMR) though one would not expect this effect to be substantial because it requires strong spin-orbit coupling. To rule out this possible artifact, we measured TAMR effect on the sample device presented in the main text. As shown in Figure S12, there is no observable TAMR signal, which rules out TAMR as the origin of the observed spin signal in the main text.

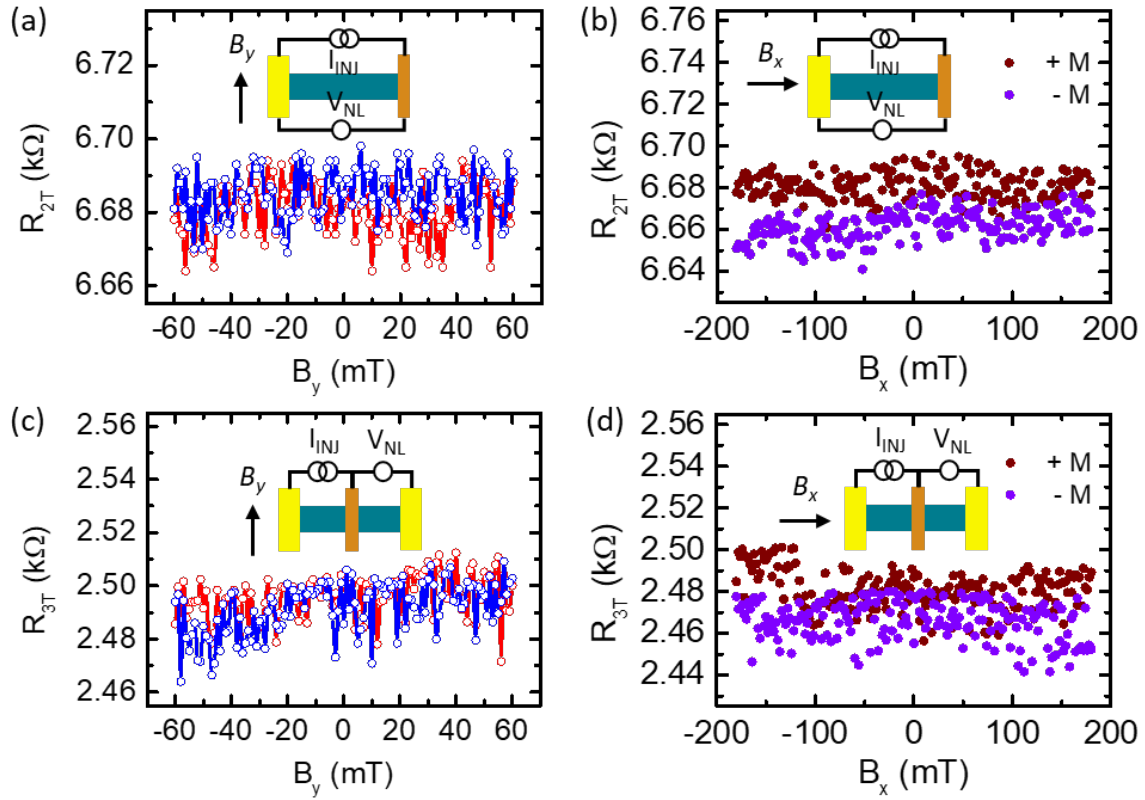

Figure S12. TAMR effect measurement at T = 20 K and $V_{gate}$ = 20 V. a. and b. are two terminal measurement, c. and d. are three terminal measurement. In a. and c. the red (blue) curve is for increasing (decreasing) magnetic field. In b. and d. the wine (violet) curve is for Co magnetization along +y (-y) direction. The insets are the measurements set-up. There is no observable TAMR effect.